\documentclass[aps,showpacs,superscriptaddress,twocolumn]{revtex4}

\usepackage{amsfonts}
\usepackage{amsmath}
\usepackage{amsthm}
\usepackage{amscd}
\usepackage{amssymb}
\usepackage{subfigure}
\usepackage{amsxtra}
\usepackage{bm}           
\usepackage{bbm}
\usepackage{exscale}
\usepackage{epic,rotating}
\usepackage{graphics,color,dsfont}
\usepackage{latexsym}
\usepackage{enumerate}
\usepackage{dcolumn}      
\usepackage{pstricks,pst-grad,fancybox,graphics}
\usepackage[dvips]{epsfig}

\def\bi#1\ei {\begin{itemize}#1\end{itemize}}
\def\bn#1\en {\begin{enumerate}#1\end{enumerate}}
\def\bea#1\eea {\begin{align}#1\end{align}}
\def\bean#1\eean {\begin{align*}#1\end{align*}}
\def\ben#1\een {\begin{equation*}#1\end{equation*}}
\def\be#1\ee {\begin{equation}#1\end{equation}}
\def\bes#1\ees {\begin{equation}\begin{split}#1\end{split}\end{equation}}
\def\bear#1\eear {\begin{eqnarray}#1\end{eqnarray}}
\def\bear#1\eear {\begin{eqnarray*}#1\end{eqnarray*}}

\newcommand{\ket}[1]{\ensuremath{|#1\rangle}}

\newcommand{\ketbra}[1]{\ensuremath{| #1 \rangle \langle #1 |}}
\newcommand{\mean}[1]{\ensuremath{\langle #1 \rangle}}

\newcommand{\eins}{\ensuremath{\mathbbm 1}}
\renewcommand{\qed}{\ensuremath{\hfill \blacksquare}\medskip}
\newcommand{\BB}{\ensuremath{\mathcal{B}}}
\newcommand{\HH}{\ensuremath{\mathcal{H}}}

\newcommand{\MM}{\ensuremath{\mathcal{M}}}

\newcommand{\OO}{\ensuremath{\mathcal{O}}}

\newcommand{\tr}[1]{\ensuremath{\mbox{Tr}\left( #1 \right)}}

\newcommand{\RR}{\ensuremath{\mathbbm{R}}}

\newtheorem{thm}{Theorem}

\newtheorem{prop}[thm]{Proposition}
\newtheorem{defn}[thm]{Definition}

\newtheorem{obs}[thm]{Observation}

\begin{document}

\title{Accessible nonlinear entanglement witnesses}

\author{Juan Miguel Arrazola, Oleg Gittsovich, Norbert L\"utkenhaus}
\affiliation{Institute for Quantum Computing,
University of Waterloo, 200 University Avenue West,
N2L 3G1 Waterloo, Ontario, Canada}

\affiliation{Department of Physics and Astronomy,
University of Waterloo, 200 University Avenue West,
N2L 3G1 Waterloo, Ontario, Canada}

\begin{abstract}
Verification of entanglement is an important tool to characterize sources
and devices for use in quantum computing and communication applications.
In a vast majority of experiments entanglement witnesses (EW) are used in order to prove
the presence of entanglement in a quantum state. EWs can be constructed from available measurement
results and do not require a reconstruction of the whole density matrix (full
tomography), which is especially valuable for high-dimensional systems. We provide a method to construct {\it accessible nonlinear EWs}, which
incorporate two important properties. First, they improve on linear EWs in the sense that each non-linear EW detects
more entangled states than its linear counterpart and therefore allow the verification of entanglement without
critical dependence on having found the 'right' linear witness.
Second, they can be evaluated using exactly the same data as for
the evaluation of the original linear witness. This allows a reanalysis of
published experimental data to strengthen statements about entanglement
verification without the requirement to perform additional measurements. These particular properties
make the accessible nonlinear EWs attractive for the implementations in current experiments, for they
can also enhance the statistical significance of the entanglement verification.
\end{abstract}


\date{\today}

\maketitle

Entangled states are an important resource for performing quantum information processing tasks \cite{guhne09a}. As a tool
for characterization of this resource entanglement
verification finds broad application in theoretical and practical testing of quantum resources and devices, such as quantum
memories, quantum repeaters or quantum channels \cite{Killoran10a,killoran10b}, which are essential for the creation,
storage and distribution of quantum information.

In experimentally relevant scenarios entanglement witnesses (EW) are playing a central role as a tool for entanglement detection
\cite{guhne09a,RevModPhys.81.865}. Recently, a theoretical construction has been suggested which extends  witnesses that
use linear combinations of observed expectation values to those using nonlinear
functions \cite{guehne_nonlinear_2006,guehne_nonlinear_2007,moroder_iterations_2008}. The nonlinear extension of a linear EW is stronger in the
sense that it detects a strictly larger set of entangled states. However, the evaluation of a generic nonlinear extension will typically
require additional measurements compared to the evaluation of the linear EW. This fact can be a drawback  for experimental applications.

In all experiments for entanglement detection which use EWs as a tool, the observable corresponding to the witness is
not measured directly. Instead, the EW observable is decomposed in terms of local observables. Then, one performs a series of measurements corresponding to the measurement of each term in the decomposition.
Such a decomposition is not unique and can be chosen according to the specific nature of the experiment \cite{guhne09a}.

The problem of entanglement detection with a fixed restricted set of measurements has been already addressed in Ref. \cite{curty05a}.
Investigation of this problem naturally leads to the definition of {\em verifiable states}.
For the given set of restricted measurements a state is called verifiable if the outcomes of the measurements are not compatible with the outcomes
of the same measurements on any separable state. The set of all linear witnesses which can be constructed from this restricted set of measurements
is called the {\em verification set}. Characterizing the verification set
or finding a witness for a given verifiable state is a challenging problem \cite{curty05a}, which is in fact
as hard as finding an EW for an arbitrary entangled state \cite{RevModPhys.81.865}.

Here we introduce a framework for the entanglement detection when only a restricted set of local observables is measured and no further knowledge about
the state can be obtained. It leads to a family of easily computable entanglement criteria - {\it accessible nonlinear EWs}.
Accessibility of a nonlinear witness (for which a rigorous definition is given later) refers to the fact that it
can be written as a rational function of expectation values that can be accessed by just performing the restricted set of
measurements,
i.e. the same measurements that are necessary for the evaluation of the linear EW. As the range of detected states is significantly extended for non-linear witnesses, we can to some degree avoid the problem of
searching  through the verification set of linear witnesses. This comes at the price that not all verifiable states will be
detected.
Furthermore we address the question under which conditions an accessible nonlinear EW can be constructed.
In a general scenario we provide sufficient conditions for the accessibility of the nonlinear extension.
For several relevant but more specific scenarios we find necessary conditions as well.
Besides, accessible nonlinear EWs turn out to have higher detection significance when
evaluated from the experimental data.

{\it Main idea.-} Let us first introduce some notation. Denote by $\HH_A$ ($\HH_B$) Hilbert space of
dimension $d_A$ ($d_B$) and by $\BB(\HH_A)$ ($\BB(\HH_B)$) the set of all bounded linear operators
on the corresponding Hilbert space. Quantum states on a bipartite Hilbert space $\HH_{AB}$ are described by
positive semi-definite operators with unit trace and denoted by $\rho_{AB}$. A linear EW on $\HH_{AB}$ is
a Hermitian operator $W_{AB}$, such that
\bea
&\forall \rho^s_{AB} \mbox{ - separable, } \tr{W_{AB}\rho^s_{AB}}\geq 0\nonumber\\
& \exists \rho_{AB} \mbox{ - entangled, s.t. } \tr{W_{AB}\rho_{AB}}< 0
\eea

Initial work on non-linear witnesses \cite{guehne_nonlinear_2006} concentrated on extension of linear witnesses that were equivalent to some negative partial transposition criterion.  Subsequently, it was pointed out that by employing Choi-Jamio{\l}kowski isomorphism
\cite{jamiolkowski72a,lewenstein_optimization_2000} it is in principle possible to construct nonlinear improvements
for any given linear EW \cite{guehne_entanglement_2006}. We will review nonlinear witnesses using this generalized view.
Any entanglement witness
$W_{AB}\in\BB(\HH_A\otimes\HH_B)$ is equivalent to a positive but not completely positive (PnCP) map
$\Lambda_{W} :\BB(\HH_A)\rightarrow \BB(\HH_B)$ via the following relationship
\be
\Lambda_{W}(X_A) = \mbox{Tr}_A\left(W_{AB}^{T_A}\; \;X_A\otimes\eins_B\right)\in\BB(\HH_B),
\ee
for any $X_A\in\BB(\HH_A)$, where the partial transposition is taken w.r.t. an orthonormal
basis $\{\ket{k}_{k=1}^{d_A}\}$.
Due to the isomorphism of the map $W_{AB}\mapsto\Lambda_{W}(.)$ the witness is restored by applying an extension of $\Lambda_{W}$ to
the projector onto the maximally entangled state on $\HH_{AA'}$, where $\HH_{A'}\cong \HH_A$,
\be
\eins_A\otimes\Lambda_{W}[\mbox{P}_{\Psi_{AA'}^+}]:=\tilde{\Lambda}_{W}[\mbox{P}_{\Psi_{AA'}^+}]=W_{AB}.
\ee
In the remainder of the paper we will omit the subscript $AB$ of $W$, unless we state explicitly to which two parties we refer to,
while keeping in mind that $W$ always acts on a bipartite Hilbert
space $\HH_{AB}$ and the extension of the map $\tilde{\Lambda}$, corresponding to $W$, acts always from $\HH_{AA'}$ to $\HH_{AB}$.

In order to derive a nonlinear EW we first note that $\tr{\rho \OO\OO^{\dagger}}$ is positive semi-definite for any state $\rho$ and
any operator $\OO$. As a consequence, for any decomposition $\OO=\sum_{i=1}^n c_i\OO_i$ with free parameters $c_i$ and any set of operators $\OO_i$, this is equivalent to the positive semi-definiteness of a matrix $M_{\rho}=\left\{\tr{\rho \OO_i\OO^{\dagger}_j}\right\}_{i,j=1}^n$.
Then, by choosing $\OO_1$ as the projector onto the maximally entangled state $\mbox{P}_{\Psi^+}$
and $\OO_2$ as some unitary $U$ we see that for all separable bipartite states
$\rho_s\in\BB(\HH_{AB})$ the matrix
\bea
&M(\rho_s) = \nonumber\\
&\left(\begin{array}{cc}
\tr{\tilde{\Lambda}^{\dagger}_{W}[\rho_s]\eins} & \tr{\tilde{\Lambda}^{\dagger}_{W}[\rho_s]\mbox{P}_{\Psi^+}U}\\
\tr{\tilde{\Lambda}^{\dagger}_{W}[\rho_s]U\mbox{P}_{\Psi^+}}^* & \tr{\tilde{\Lambda}^{\dagger}_{W}[\rho_s]\mbox{P}_{\Psi^+}}
\end{array}\right)\label{tobisiteration}
\eea
must be positive semi-definite \cite{moroder_iterations_2008}. Here we introduced $\tilde{\Lambda}^{\dagger}$ as the adjoint map of $\tilde{\Lambda}$.
Since, $\Lambda_W$ is a PnCP map $\tilde{\Lambda}_W$ will transform separable states into positive operators, and so
will its adjoint $\tilde{\Lambda}^{\dagger}$.
Therefore for general bi-partite states $\rho$,
failure of the matrix in Eq. (\ref{tobisiteration}) to be positive is a conclusive proof that that particular $\rho$ must be
entangled. By using the definition of the adjoint map again, while noting that
$\tilde{\Lambda}_{W}[\mbox{P}_{\Psi_{AA'}^+}]=W_{AB}$ and assuming that $\tr{\rho\tilde{\Lambda}_W[\eins]}\neq 0$, we construct from the determinant of  Eq. (\ref{tobisiteration}) a nonlinear function which improves the entanglement detection given by the linear EW,
\be
w_{NL}(\rho) = \tr{\rho W} - \frac{|\tr{\rho\tilde{\Lambda}_{W}[\mbox{P}_{\Psi^+}U]}|^2}{\tr{\rho\tilde{\Lambda}_W[\eins]}}.
\label{NLindeaeq}
\ee
That is, if for some $U$ we have that $w_{NL}(\rho)<0$, then the state $\rho\in\BB(\HH_{AB})$ must be entangled.  The new criterion \eqref{NLindeaeq} detects more entangled states, for we subtract a strictly positive number from the expectation value of $W$.

The quantity $w_{NL}(\rho)$ can be represented as an expectation value of an operator $W_{NL}$, which we call a nonlinear
EW. From Eq. (\ref{NLindeaeq}) it is straightforward to verify that $W_{NL}$ is given by
\bea
W_{NL}&=\tilde{\Lambda}_W\left[Q_{NL}Q^{\dagger}_{NL}\right]\nonumber\\
Q_{NL}&=\mbox{P}_{\Psi^+}U - \varkappa\tr{\rho\tilde{\Lambda}_W[\mbox{P}_{\Psi^+}U]}\eins\\
\varkappa^{-1}&=\tr{\rho\tilde{\Lambda}_W[\eins]}.\nonumber
\eea
The case of $\varkappa\neq 1$ reveals the fact that $\Lambda_W$
can be non-unital, meaning that $\tilde{\Lambda}_W$ does not necessarily map the identity onto the identity.

Let us mention some facts about the nonlinear EWs. To begin with, note that Eq. (\ref{NLindeaeq}) can be easily
iterated in a similar way as it is done in Ref. \cite{moroder_iterations_2008}. There, for example, nonlinear
improvements for all {\it fully PPT witnesses} \cite{jungnitsch_taming_2011,jungnitsch_entanglement_2011},
i.e. EWs of the form $W=P^{\Gamma}$, with $P\geq 0$ and $\Gamma$ the partial transposition w.r.t. either of parties,
were considered. The iteration process in the general case is established by defining
\bea
Q_1&=\mbox{P}_{\Psi^+}U_0 - \varkappa\tr{\rho\tilde{\Lambda}_W[\mbox{P}_{\Psi^+}U_0]}\eins,\label{Qndefn}\\
Q_{n}&=Q_{n-1}U_{n-1} - \varkappa\tr{\rho\tilde{\Lambda}_W[Q_{n-1}U_{n-1}]}\eins,\: n\geq 2.\nonumber
\eea
Then the $n$th nonlinear improvement is given by $W_n:=\tilde{\Lambda}_W[Q_nQ_n^{\dagger}]$. Moreover, a simple calculation
shows that $w_n(\rho):=\tr{\rho W_{NL}}$ is given by a recurrence relation
\be
w_n(\rho)=w_{n-1}(\rho)-\varkappa c_{n-1}(\rho),\label{recurrrel}
\ee
with
\be
c_{n-1}(\rho):=|\tr{\rho\tilde{\Lambda}_W[Q_{n-1}U_{n-1}]}|^2 \; .
\ee
If $w_n(\rho)<0$ for some $n$ then $\rho$ must have been
entangled.

The evaluation of the quantities $w_n(\rho)$ in general would require the knowledge of the whole density
matrix. In order to overcome this restrictive circumstance, we can employ the aforementioned fact that the linear witnesses
are not measured directly but rather decomposed in a linear combination of local measurable observables which are
then measured. Therefore the starting point of all our studies will be always a particular decomposition of a linear EW
in terms of the local observables $A_i\in \BB(\HH_A)$ and $B_i\in \BB(\HH_B)$ on Alice's and Bob's side.
Let us write this decomposition as $W=\sum_{i=1}^N c_i A_i\otimes B_i$.
By measuring the observables $\{A_i\otimes B_i\}_{i=1}^N$ we naturally induce
a linear and, in general, not injective map $\MM : \rho\mapsto \MM(\rho)\in \RR^N$,
where the space $\RR^N$ consists of the expectation values of observables $\{A_i\otimes B_i\}_{i=1}^N$ in
all possible physical states $\rho\in\BB(\HH_{AB})$ \cite{curty05a}. The map $\MM$ maps a convex set
of separable states $S$ onto a convex set $S'=\MM(S)\in \RR^N$ - the equivalence class of all separable
states that are compatible with the expectation values $\{\mean{A_i\otimes B_i}\}_{i=1}^N$. The set of all linear EWs
$W=\sum_{i=1}^N \alpha^W_i A_i\otimes B_i$ for different coefficients $\{\alpha_i^W\}$ form the verification set
and are equivalent to a set of hyper-planes in $\RR^N$, each dividing $\RR^N$ into two halves, one of which
contains the set $S'$ \cite{curty05a}.

It is evident now that no additional measurements are needed if $w_n(\rho)$ can be evaluated from the expectation values $\{\mean{A_i\otimes B_i}\}_{i=1}^N$ only and, therefore, defines a hyper-surface in $\RR^N$.
This leads us to the definition of an accessible nonlinear EW as a nonlinear improvement of a linear EW from the verification set.

\begin{defn}(Accessible nonlinear EWs)
For a given decomposition of a linear EW $W=\sum_{i=1}^N c_i A_i\otimes B_i$ in terms of locally measurable observables $A_i\otimes B_i$, a nonlinear EW $W_{n}=\tilde{\Lambda}_W[Q_nQ_n^{\dagger}]$ is accessible if it defines a rational function in the space $\RR^N$ of expectation values $\{\mean{A_i\otimes B_i}_{\rho}\}_{i=1}^N$.
\end{defn}

As we will show now, it is possible to formulate conditions when the knowledge
of the set of expectation values $\{\mean{A_i\otimes B_i}\}_{i=1}^N$ is sufficient (and in some cases even
necessary and sufficient) for the calculation of $w_n(\rho)=\tr{W_n\rho}$.


Our first goal is to find the conditions when, for the first step of the iteration, $w_{NL}(\rho)$ in Eq. (\ref{NLindeaeq})  can be written
as a rational function in the space $\RR^N$ that is defined by the restricted set of measurements. For that it is sufficient to show that the operators
$\tilde{\Lambda}_W[\mbox{P}_{\Psi^+}U]$ and $\tilde{\Lambda}_W[\eins]$ can be decomposed only in terms of the observables
$\{A_i\otimes B_i\}_{i=1}^N$. This will imply that the quadratic term in Eq. (\ref{NLindeaeq}) can be
computed from the expectation values $\mean{A_i\otimes B_i}_{\rho}$ and no further measurements are needed
or equivalently that $w_{NL}(\rho)$ is a rational function in $\RR^N$.

Formally, the sufficient condition from the previous paragraph can be grasped by the following Observation.
\begin{obs}{\bf (Sufficient conditions for accessible nonlinear EW)}\label{suffcond} Let $W=\sum_ic_i A_i\otimes B_i$
be a decomposition of a linear EW $W$ in terms of local observables $A_i$ and $B_i$.
Denote by $V=\mbox{span}\{A_i\otimes B_i\}_{i=1}^N$ the linear subspace of $\BB(\HH_{AB})$ containing all
observables in the decomposition of the linear witness $W$ and by $V'$ the subspace of the domain of $\tilde{\Lambda}_W$,
such that $\tilde{\Lambda}_W[V']\subseteq V$.
Then $w_{NL}(\rho)$ is an accessible nonlinear EW if
\begin{itemize}
\item[(i)] The elements of $V'$ form an algebra with respect to the canonical matrix multiplication which is
associative and contains the identity (i.e. form a unital associative algebra)
\item[(ii)] $U\in V'$.
\end{itemize}
\end{obs}
{\it Proof:} First note that $\mbox{P}_{\Psi^+}\in V'$,
since $\tilde{\Lambda}_W[\mbox{P}_{\Psi^+}]=W\in V$. As $V'$ is an unital algebra
and $U\in V'$ it follows that $\eins,\mbox{P}_{\Psi^+}U\in V'$ and therefore $\tilde{\Lambda}_W[\eins]$,
$\tilde{\Lambda}_W[\mbox{P}_{\Psi^+}U]\in V$. This implies that the operator
$\tilde{\Lambda}_W[\mbox{P}_{\Psi^+}U]$ can be decomposed in terms of local
observables $\{A_i\otimes B_i\}_{i=1}^N$ only and $w_{NL}(\rho)$ is accessible.
\qed

Furthermore, if the conditions of Observation \ref{suffcond}
are fulfilled, then $w_n(\rho)$ represents an accessible nonlinear EW for any $n$.
This can be readily shown by applying the method of the mathematical induction in $n$.
Since the conditions of Observation \ref{suffcond} hold,
the definition of the operator $Q_1$ implies that it lies in the algebra $V'$ and therefore
$w_1(\rho)$ is accessible. Thus if $Q_{n-1}$ lies in $V'$ and $U_{n-1}\in V'$ so does
$Q_n$, which implies that $w_n(\rho)$ is accessible.
Formally we have:
\begin{obs} For any given decomposition of a linear EW $W=\sum_ic_i A_i\otimes B_i$,
which fulfills the conditions of Observation \ref{suffcond} there exists a sequence of accessible
nonlinear EWs $w_n(\rho)=w_{n-1}(\rho)-\varkappa c_{n-1}(\rho)$ with $\varkappa$, $w_n(\rho)$ and
$c_n(\rho)$ defined as above.
\end{obs}

Before we go over to examples of accessible nonlinear witnesses, we will now provide structures that are useful for these examples and many experimental implementations.

{\it An analytical formula for the iterations.-} For a special choice of unitary operators in the iteration process,
$\lim_{n\rightarrow\infty}w_n$ can be calculated analytically and gives a strong accessible nonlinear EW.

\begin{prop}{\bf(Analytical formula for accessible nonlinear EW)}\label{analytic}
Let $W=\sum_ic_i A_i\otimes B_i$ be a decomposition of a linear EW.
If the conditions of Observation \ref{suffcond} hold, then for constant $U = U_n$ (for all $n$) which also satisfies $U^2=\eins$ the following alternative is true
\bea
(i) \mbox{ if }\varkappa|k(\rho)|&<1,\nonumber\\
\mbox{then } w_{\infty}&(\rho):=\lim_{n\rightarrow\infty}w_n(\rho)\label{winfty}\\
=&\tr{\rho W}- \varkappa |c(\rho)|^2 - \frac{\varkappa |d(\rho)|^2 }{1-(\varkappa|k(\rho)|)^2},
\nonumber\\
&\mbox{with}\nonumber\\
&k(\rho) = \tr{\rho\tilde{\Lambda}_W[U]},\nonumber\\
&\varkappa^{-1} = \tr{\rho\tilde{\Lambda}_W[\eins]}\nonumber\\
&c(\rho) = \tr{\rho\tilde{\Lambda}_W[P_{\Psi^+}U]},\nonumber\\
&d(\rho) = \tr{\rho\tilde{\Lambda}_W[P_{\Psi^+}]}- \varkappa c(\rho)k(\rho)\nonumber\\
(ii) \mbox{ if } \varkappa|k(\rho)|&\geq 1,\; d(\rho)> 0,\nonumber\\
\mbox{then } w_n&(\rho)\mbox{ diverges to }-\infty, \nonumber
\eea
\end{prop}
{\it Proof:} In order to prove the statement first note that
by applying twice the recurrence relation Eq. (\ref{Qndefn}) to $\tr{\rho\tilde{\Lambda}_W[Q_n]}$, while
keeping in mind that $U_n^2=\eins$, we have $\tr{\rho\tilde{\Lambda}_W[Q_n]}=0$ for all $n>1$. Hence,
by using the same recurrence relation again we immediately see that the coefficients $c_n(\rho)$ form a geometric
progression $c_n(\rho)=\varkappa^2|k(\rho)|^2c_{n-1}(\rho)$. Thus
$w_n(\rho) = \tr{W\rho} - \varkappa |c(\rho)|^2- \varkappa |d(\rho)|^2\sum_{m=1}^{n-1}(\varkappa|k|)^{2(m-1)}$.
Taking the limit $n\rightarrow \infty$ proves the claim.\qed

Notably, for the special choice of the unitary as in Proposition \ref{analytic}, we can strengthen Observation
\ref{suffcond} and formulate conditions which are necessary and sufficient for the evaluation of the analytical
formula.
\begin{obs}{\bf (Necessary and sufficient conditions for the analytic formula)}
Let $W=\sum_ic_i A_i\otimes B_i$ be a decomposition of a linear EW and let us choose the unitary $U$ as in the
previous observation, $U_n = U$ and $U^2=\eins$.
Then $w_{n}(\rho)$ is an accessible nonlinear EW for all values of $n$ if and only if $k(\rho)$, $\varkappa(\rho)$ and $c(\rho)=\tr{\rho\tilde{\Lambda}[P_{\Psi^+}U]}$ are accessible.
\end{obs}

{\it Proof:} First assume that $k(\rho)$, $\varkappa(\rho)$ and $d(\rho)$ are accessible. Then $w_n(\rho)$ must be accessible
as it is a function of $k(\rho), \varkappa(\rho)$ and $d(\rho)$ only, for all values of $n$. This finishes the first part
of the proof.

Now assume every $w_n(\rho)$ is accessible. Define $v_i:=\tr{\rho A_i\otimes B_i}$ and
$z_j:=\tr{\rho Z_j}$ where the set of hermitian operators $\{Z_j\}$ form a basis of the orthogonal complement
$V^{\perp}$ of $V=\mbox{span}\{A_i\otimes B_i\}_{i=1}^N$. Then a quantity $q(\{v_i\},\{z_i\})$ is accessible if it can be expressed
as a function of the variables $v_i$ only. Equivalently one can say that all derivatives of $q$ with respect to every $z_j$
must vanish for all values of the variables $\{v_i\}$ and $\{z_j\}$.

First note that for $c(\rho)=0$, we find that $w_n(\rho)=w_0(\rho)$ for all $n$ and for any values of $\varkappa(\rho)$ and $k(\rho)$, since $c_n(\rho)=\varkappa^2|k|^2c_{n-1}(\rho)$ when $U_n^2=\eins$. In this case the nonlinear improvements coincide with the expectation value of the linear witness and therefore do not provide any advantage, so we do not have to worry about the accessibility of $\varkappa(\rho)$ and $k(\rho)$ at all. So from now on we can assume that  $c(\rho)\neq 0$ whenever necessary.

In general, we can write $c=\sum_i a_i v_i +\sum_{j} x_j z_j$ and $\varkappa^{-1}=\sum_{i} b_i v_i +\sum_{j} y_j z_j$ for some fixed numbers $\{a_i\},\{x_j\},\{b_i\},\{y_j\}$. We want to show that $x_j=y_j=0$ for all values of $j$. Using the fact that $w_1$ and $\tr{\rho W}$ are accessible together with the above expressions for $c$ and $\varkappa^{-1}$ one can show that the condition $\frac{\partial}{\partial z_l}\varkappa |c|^2=0$, for any $l=1,\ldots,dim(V^{\perp})$ is equivalent to the system of linear equations

\begin{align}
&\gamma_l |x_l|^2=0\label{eq1}\\
&y_l|x_l|^2=0\label{eq2}\\
&\beta_l\gamma_l=\alpha_l y_l\label{eq3},
\end{align}

where

\begin{align}
\alpha_l &=\sum_{i}\sum_{j} a_ia_j^*v_iv_j+\sum_{i}\sum_{j\neq l} (a_ix_j^*+a_i^*x_j)v_iz_j\label{alphal}\\
\beta_l &=\sum_{i}(a_ix_l^*+a_i^*x_l)v_i+\sum_{j\neq l} (x_jx_l^*+x_j^*x_l)z_j\label{betal}\\
\gamma_l &=\sum_{i} b_i v_i+\sum_{j\neq l} y_j z_j.
\end{align}

Recall from \eqref{NLindeaeq} that we always have $\varkappa^{-1}=\gamma_l+y_lz_l\neq 0$ and therefore Eqns. \eqref{eq1} and \eqref{eq2} imply $|x_l|=0$ (otherwise both $\gamma_l$ and $y_l$ must be equal to zero, which would imply $\varkappa^{-1}=0$).
Since this holds for any arbitrary $l$,
it proves that $d$ is accessible. Further, because of $|x_l|=0$, Eq. (\ref{betal}) becomes $\beta_l=0$ and Eq. (\ref{alphal}) becomes
$\alpha_l =\sum_{i}\sum_{j} a_ia_j^*v_iv_j=|c|^2$. Moreover Eq. \eqref{eq3} gives $\alpha_l y_l=0$.
Since $\alpha_l=|c|^2\neq 0$ it must be that $y_l=0$ for all values of $l$. This proves that $\varkappa$ is also accessible.
Finally, using the fact that $w_2$ and $w_3$ are accessible, it is straightforward to show that $\varkappa |k|$ is
accessible and this implies that $k$ is accessible as well.
\qed

{\it Examples.-} As a first example we consider an optimal decomposable witness for two qubits $W_0=P^{\Gamma}$,
where $P$ is a positive semi-definite operator and $\Gamma$ denotes the partial transposition w.r.t.
either one of the qubits \cite{augusiak_optimal_2011}. Note that transposition is a PnCP map that leads to a
necessary and sufficient criterion for entanglement detection in two qubit case \cite{peres96a}.
The decomposition of such two-qubit EW in terms of local observables has been extensively studied.
Let us consider an EW with the decomposition of the form
$W_0=(\eins + \sum_{\alpha=x,y,z}\sigma_{\alpha}\otimes\sigma_{\alpha})/4$. This is an optimal entanglement witness \cite{guhne02a,guhne03a}.
This EW does not detect the entanglement of the state $(\ket{00}+\ket{11})/\sqrt{2}$,
which is one of the Bell states. However, that entanglement is detected by the nonlinear improvement in Eq. (\ref{winfty}).
Indeed $\tilde{\Lambda}_W$ corresponds to the partial transposition and consequently $\varkappa^{-1}=1$.
Direct calculation for the choice $U=(\eins+\sum_{\alpha=\{x,y,z\}}\sigma_{\alpha}\otimes\sigma_{\alpha})/2$ which satisfies $U^2=\eins$, gives $|k|=1$,
with $d(\rho)\neq 0$ and hence $w_{\infty}(\rho)=-\infty$, meaning that the state is detected by the constructed
accessible nonlinear EW $w_{\infty}(\rho)$.

\begin{figure}
\includegraphics[width=0.99\columnwidth]{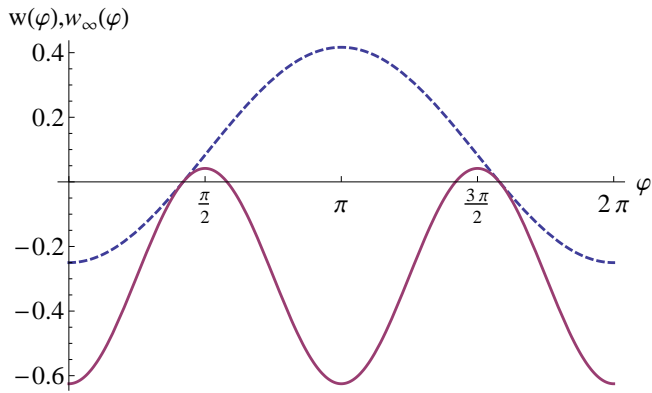}
\caption{(Color online) Value of the linear witness $W_0$ (dashed blue curve) and its nonlinear improvement $W_{\infty}$ (purple curve) for the state $\rho(\varphi)=(2/3)\ketbra{\varphi}+(\eins/12)$ where $\ket{\varphi}=\frac{1}{\sqrt{2}}(\ket{01}-e^{i\varphi}\ket{10})$. The starting witness is $W_0=(\eins + \sum_{\alpha=x,y,z}\sigma_{\alpha}\otimes\sigma_{\alpha})/4$ and the nonlinear improvement was constructed by choosing $U=\sigma_z\otimes\sigma_z$.
\label{phase}}
\end{figure}

In Fig. \ref{phase} we illustrate the results for a particular family of the states $\rho(\varphi)=(2/3)\ketbra{\varphi}+(1/12)\eins$, with $\ket{\varphi}=\frac{1}{\sqrt{2}}(\ket{01}-e^{i\varphi}\ket{10})$ with
$\varphi \in [0,2\pi]$. The value of the white noise admixture has been chosen to demonstrate a typical behavior. As we see, the linear witness detects entanglement only outside some interval containing $[\pi/2,3\pi/2]$, while the accessible nonlinear
extension detects most entanglement within the interval. Note that the state is entangled for all values of $\varphi$, so this
particular nonlinear extension does not detect all entanglement.

This example provides a nice demonstration of the general fact that the accessible nonlinear EWs always detect more states than their linear ancestors.

As our second example we consider improvements for a particular
decomposition of a witness that detects bound entanglement in the four qubit Smolin state
$\rho_S = \frac{1}{4}\sum_{k=0}^3 \ketbra{\Psi_k}_{12}\otimes\ketbra{\Psi_k}_{34}$
\cite{smolin_four-party_2001}. Notably this state was recently
prepared in the lab and its bound entanglement was confirmed experimentally \cite{lavoie_experimental_2010}.
The witness $W$ with the decomposition $W = \frac{1}{16}(\eins - \sum_{\alpha = x,y,z} \sigma_{\alpha}^{\otimes 4})$
is optimal for this state \cite{pittenger_convexity_2001}.

\begin{figure}
\includegraphics[width=0.99\columnwidth]{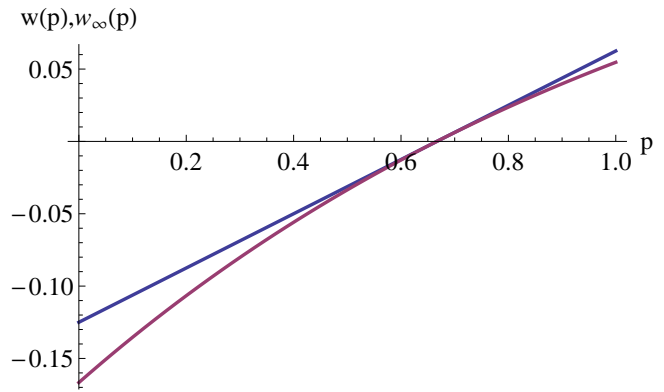}
\caption{(Color online) Value of the
linear witness $W_0$ (blue line) and of its nonlinear improvement $w_{\infty}(\rho)$ (purple curve) as a
function of noise parameter $p$ for the Smolin state mixed with the white noise $\rho_S(p)=(1-p)\rho_S+p\eins/16$.
$w_{\infty}(\rho)$ is calculated according to Eq. (\ref{winfty}) with
$U=\frac{1}{4}(\eins + \sum_{\alpha = x,y,z} \sigma_{\alpha}\otimes\sigma_{\alpha})$
\label{NLSmolinAnalytic}}
\end{figure}

An accessible nonlinear witness can be derived from this witness decomposition by choosing the unitaries $U_n=(\eins + \sigma_x\otimes\sigma_x + \sigma_y\otimes\sigma_y +
\sigma_z\otimes\sigma_z)/2$ for all $n$ and constructed the nonlinear EW according to Eq. (\ref{winfty}).
In Fig. \ref{NLSmolinAnalytic} we present a numerical calculation for the detection of the Smolin state mixed with the white noise
$\rho_S(p)=(1-p)\rho_S+p\eins/16$ depending on the noise parameter $p$. The expectation value of the linear
EW is represented by the line $w(p)$ and its improvement by the curve $w_{\infty}(p)$. Both witnesses tolerate
a critical amount of noise of $p=2/3$ as the linear witness happens to be optimal at that point. Other types of noise would reveal the advantage of the nonlinear improvement in a manner similar to the example in Fig. \ref{phase}.


{\it Volume of detectable states and detection significance of the accessible nonlinear EWs.-}

For the entanglement verification in experiments, detection significance is an important issue, which has to be addressed.
The common and widely used approach to put error bars on the measured data - the Gaussian error propagation
(see e.g. \cite{EadieBook}), has recently lead to somewhat counterintuitive statements, stating, for example, that
the smaller negative value of a linear entanglement witness does not necessarily imply higher detection significance
\cite{jungnitsch_increasing_2010}.

In order to make reliable and meaningful statements for detection significance in entanglement verification experiments
one ought to apply a more consistent framework. Such a framework has been recently presented by M. Christandl and R. Renner in
Ref. \cite{christandl_reliable_2011}. There, the outcomes of $n$ runs of an experiment leads to an estimate density
$\mu_{n}(\rho)$, which can be seen as a measure on the space of all states.
According to Chapter 6 in Ref. \cite{christandl_reliable_2011} we assign the probability for a state to be in
a subset $\Gamma$ of the state space w.r.t. the estimate density $\mu_{n}(\rho)$ as follows:
\be
\int_{\Gamma}\mu_{n}(\rho)d\rho = P_{\mu_n}(\Gamma).
\ee

This picture can be related to the detection significance provided by the entanglement witnesses in the following sense. Denote
by $\Gamma_W$ the set of all states that are detected by a witness $W$. According to Christandl-Renner framework any measured
state will give rise to an estimate density $\mu_n$ which is typically non-zero on the whole state space.
Then the detection significance
of the state is given by $P_{\mu_n}(\Gamma_W)$. Moreover, we say that a state is detected with a high significance
if $P_{\mu_n}(\Gamma_W)=1-\epsilon_W$, where $\epsilon_W$ is small comparing to 1.

From the previous considerations and from our first example it is clear that the volume of the states detected by the accessible
nonlinear EW always includes the volume of the states detected by its linear ancestor. Interestingly, analyzed
according to the framework of Ref. \cite{christandl_reliable_2011} the detection
significance can only increase for accessible nonlinear EWs. Indeed, denote by $\Gamma_W$ and $\Gamma_{W_{NL}}$ the sets
of states that are detected by the linear and the accessible nonlinear EW respectively. Then, since $\Gamma_W$
is strictly contained in $\Gamma_{W_{NL}}$, for any state we have
\be
1-\epsilon_W = P_{\mu_{n}}(\Gamma_W) < P_{\mu_{n}}(\Gamma_{W_{NL}}) =1- \epsilon_{W_{NL}}
\ee
and the detection efficiency of the accessible nonlinear EW is strictly bigger than the detection efficiency of the linear EW.

{\it Possible extensions and outlook.-}
The theory developed in this paper may be applied to a wide range of linear EWs and various examples
of accessible nonlinear EWs can be constructed. One of the open questions which is worth investigating
is whether for a given decomposition of a linear EW there exists an accessible non-linear extension.
Moreover, it is interesting to understand
if it is possible or not to construct nonlinear improvements which would detecting genuine multipartite entanglement.

{\it Conclusion.-} In this paper we investigated the separability
of quantum states with only a restricted set of measurements using accessible nonlinear
EWs. We have demonstrated that accessible nonlinear EWs detect entanglement in
many states, which are not detected by the initial EW.

Moreover, we provided error analysis for the accessible nonlinear EWs and showed that the detection
is performed with high statistical significance.

Therefore our framework provides a new systematic approach for studying entanglement of bipartite
quantum systems both in theory and in experiment.

{\it Acknowledgements.-} We thank T.Moroder, O. G{\"u}hne and M. Piani for fruitful discussions. This
work has been supported by the Industry Canada and NSERC Strategic Project Grant
(SPG) FREQUENCY.


\bibliography{NLWrefs,../../BibTex/qit_20120127}
\bibliographystyle{../../BibTex/myapsrev}

\end{document}